\documentclass[epj]{svjour}
\usepackage{latexsym}
\usepackage{graphics}
\usepackage{amsmath}
\usepackage{color}
\usepackage{graphicx}
\usepackage[normalem]{ulem}

\usepackage[english]{babel}
\usepackage{fancyref}
\usepackage[numbers,sort&compress]{natbib}
\usepackage[separate-uncertainty=true]{siunitx}
\usepackage{hyperref}
\hypersetup{
	colorlinks   = true,    
	urlcolor     = blue,    
	linkcolor    = blue,    
	citecolor    = blue     
}

\usepackage{caption}

\begin{document}

\title{
Reactive molecular dynamics simulations of organometallic  compound W(CO)$_6$ fragmentation
}

\author{
Pablo de Vera\inst{1,}
\thanks{\email{devera@mbnexplorer.com}}
\and
Alexey Verkhovtsev\inst{1,2,3,}
\thanks{\email{verkhovtsev@mbnexplorer.com}}
\and
Gennady Sushko\inst{1}
\and
Andrey V. Solov'yov\inst{1,3}
}

\institute{
MBN Research Center, Altenh\"oferallee 3, 60438 Frankfurt am Main, Germany
\and 
Department of Medical Physics in Radiation Oncology, German Cancer Research Center, Im Neuenheimer Feld 280, 69120 Heidelberg, Germany
\and
On leave from Ioffe Institute, Politekhnicheskaya 26, 194021 St. Petersburg, Russia
}

\date{\today}

\abstract{
Irradiation- and collision-induced fragmentation studies provide information about geometry, electronic properties and interactions between structural units of various molecular systems. 
Such knowledge brings insights into irradiation-driven chemistry of molecular systems which is exploited in different technological applications.
An accurate atomistic-level simulation of irradiation-driven chemistry requires reliable models of molecular fragmentation which can be verified against mass spectrometry experiments. 
In this work fragmentation of a tungsten hexacarbonyl, W(CO)$_6$, molecule is studied by means of reactive molecular dynamics simulations. 
The quantitatively correct fragmentation picture including different fragmentation channels is reproduced. 
We show that distribution of the deposited energy over all degrees of freedom of the parent molecule leads to thermal evaporation of CO groups and the formation of W(CO)$_n^+$ ($n = 0-5$) fragments. 
Another type of fragments, WC(CO)$_n^+$ ($n = 0-4$), is produced due to cleavage of a C--O bond as a result of the localized energy deposition. Calculated fragment appearance energies are in good agreement with experimental data. 
These fragmentation mechanisms have a general physical nature and should take place in radiation-induced fragmentation of different molecular and biomolecular systems.
}

\authorrunning{de Vera, Verkhovtsev, Sushko and Solov'yov}
\titlerunning{
Reactive MD simulations of W(CO)$_6$ fragmentation
}

\maketitle

\section{Introduction}

Different radiation sources (electron and ion beams, lasers, synchrotrons) are commonly used to explore structure and dynamics of complex molecular systems \cite{ISACC_LatestAdvances_2008, RadDamage_Gustavo, SynchrRad_StructBiol}.
Upon interaction with radiation a molecular target becomes excited or ionized and may relax via different dissociation mechanisms \cite{Schlatholter_2005_NIMB.233.62, Spezia_2016_FarDisc.195.599}.
The analysis of fragments produced in a collision and their kinematic properties provide useful information on the parent molecule, e.g., its 
geometry and electronic structure, the type of interaction between structural units of the system, etc. \cite{Poully_2015_PCCP.17.7172}. 
Thus, it is important to understand how the energy deposited into the target is distributed among its degrees of freedom and how this distribution leads to the observed fragmentation patterns.

Irradiation of molecular targets and subsequent irradia\-tion-driven chemistry (IDC) are exploited in modern and emerging technologies. For instance, they play an important role in modern radiotherapies such as ion-beam cancer therapy \cite{RadDamage_Gustavo, Nano-IBCT_book} that exploits the ability of heavy charged particles to inactivate living cells due to induction of complex DNA damage. Radiation is utilized also for controlling various physical and chemical processes for specific technological needs, e.g. for the formation and controllable growth of metal nanostructures under the exposure to focused electron or ion beams \cite{Utke_2008_JVacSciTechnolB.26.1197, Utke_book}, or for the production of thin films with tailored structural properties. 
%

\begin{sloppypar}
One of the technological applications of IDC is focused electron beam induced deposition (FEBID) -- a novel and actively developing nanofabrication technique that allows controllable creation of metal nano\-structures with nanometer resolution \cite{Huth_2018_review, deTeresa_2016_JPD.49.243003, Utke_book}.
FEBID is based on the irradiation of organometallic precursor molecules by 
keV electron beams 
whilst they are being deposited on a substrate. Electron-induced decomposition of the molecules releases its metallic component which forms a deposit on the surface with size similar to that of the incident electron beam (typically a few nanometers).
\end{sloppypar}

To date, a popular class of precursors for FEBID is metal carbonyls Me$_n$(CO)$_m$ \cite{Huth_2012_BeilsteinJNanotech, Kumar_2018_Beilstein.9.555}, which are composed of one or several metal atoms (Me) bound to a number of carbon monoxide ligands. 
Although such precursor molecules (for instance, W(CO)$_6$, Fe(CO)$_5$ or Co$_2$(CO)$_8$) were commonly adopted from the chemical vapor deposition  method and thus were not specifically designed to be efficiently and completely dissociated under electron irradiation, recent developments in optimization of the precursor deposition processes as well as the design of novel precursors for FEBID (e.g., HFeCo$_3$(CO)$_{12}$ \cite{Porrati_2015_Nanotechnology.26.475701}) have made it possible to fabricate fully metallic structures made of Au, Pt, Fe, Co, Pb, and Co$_3$Fe alloy \cite{Huth_2018_review}. 

Metal carbonyls have been widely studied experimentally and a substantial amount of data on thermal decomposition and electron-induced fragmentation has been accumulated over the past decades 
\cite{Beranova_1994_JAMS.5.1093, Clements_1976_MetTransB.7.171, Cooks_1990_JASMS.1.16, Wysocki_1987_IJMSIP.75.181}. 
%
This is due to their peculiar structure containing strong C--O bonds and relatively weak Me--C bonds. While the former are difficult to cleave, the latter dissociate easily, what usually happens through a sequential loss of CO groups when sufficient internal energy is available. This allows to determine internal energy distributions after electron-impact ionization, which are commonly used in mass spectrometry. Due to these properties, W(CO)$_6$ and other metal carbonyls are commonly referred to as ``thermometer molecules'' \cite{Susic_1992_JMassSpectrom.27.769, Wysocki_1987_IJMSIP.75.181}.

In recent experimental studies of W(CO)$_6$ and Fe(CO)$_5$ molecules \cite{Wnorowski_2012_RapCommunMassSpectr.26.2093, Allan_2018_PCCP.20.11692, Wnorowski_2012_IJMS.314.42, Neustetter_2016_JCP.145.054301, Lacko_2015_EPJD.69.84, Lengyel_2016_JPCC.120.17810} fragmentation mass-spectra upon electron impact were recorded as well as appearance energies for the formation of different molecular fragments were evaluated. These studies considered both low-energy electron collisions (with the typical incident energy of several eV) where fragmentation is deemed to occur via dissociative electron attachment (DEA)  \cite{Wnorowski_2012_RapCommunMassSpectr.26.2093, Allan_2018_PCCP.20.11692} and collisions with more energetic electrons 
(with the energy of one to several tens of eV) which decompose the parent molecules via dissociative ionization (DI) \cite{Wnorowski_2012_IJMS.314.42, Neustetter_2016_JCP.145.054301, Lacko_2015_EPJD.69.84, Lengyel_2016_JPCC.120.17810}. 

DI experiments with W(CO)$_6$ \cite{Wnorowski_2012_IJMS.314.42} revealed
that appearance energy for W(CO)$_5^+$ (i.e. loss of one CO group) is about 10~eV while the appearance energy for W$^+$ (loss of all ligands) is about  20~eV. The intermediate fragments, W(CO)$_n^+$ ($n = 1-4$), were also detected within this energy range. 
Another type of fragments, WC(CO)$_n^+$ ($n = 0-3$), appears above 20~eV. Furthermore, the doubly charged fragments, W(CO)$_n^{2+}$ and WC(CO)$_n^{2+}$, were detected above 40~eV, but their abundance is about an order of magnitude smaller than for the respective singly-charged species \cite{Wnorowski_2012_IJMS.314.42}.


A detailed atomistic-level understanding of IDC processes (i.e., bond dissociation and further reactivity) in molecular systems is achievable by means of computational modelling.
A rigorous quantum-mechanical description of these processes
is possible only for relatively small systems containing, at most, a few hundred atoms and evolving on the sub-picosecond time scale. Classical molecular dynamics (MD) is considered as an alternative modelling framework for much larger systems and time scales. However, standard classical MD is unable to simulate the IDC processes as it typically does not account for coupling of the system to the incident radiation nor does it describe the induced quantum transformations. 

These deficiencies were overcome recently by means of novel methodologies -- reactive CHARMM force field \cite{Sushko2015} and Irradiation Driven Molecular Dynamics (IDMD) \cite{Sushko2016, MBNbook_Springer2017} -- allowing, for the first time, high accuracy simulation of IDC in complex molecular systems. 
These methods were implemented into MBN Explorer \cite{Solovyov2012} -- a multi-purpose software package for advanced multiscale simulations of structure and dynamics of complex molecular systems with the sizes ranging from the atomic up to the mesoscopic scales \cite{MBNbook_Springer2017}. 
Within the IDMD framework the IDC transformations are treated as random, fast and local processes which can be incorporated locally into classical MD force fields 
according to the probabilities of the quantum processes that may occur in the system. Such random events result in electronic excitations involving molecular orbitals localized on a specific bond or in a specific part of a molecule. 
This eventually leads to the cleavage of particular bonds and creation of active species which can undergo further reactivity. In Ref.~\cite{Sushko2016} IDMD was successfully applied for the simulation of the FEBID process of W(CO)$_6$ precursors. It was demonstrated that IDMD is capable of reproducing experimental observations and making predictions about the morphology, molecular composition and growth rate of tungsten nanostructures emerging on a surface during FEBID \cite{Sushko2016}. These methodologies have also been applied recently to study IDC in biomolecular systems in relation to ion-beam cancer therapy \cite{deVera_2018_EPJD.72.147}.

An accurate atomistic-level simulation of IDC calls for reliable models of radiation-induced fragmentation of molecular systems. 
With respect to the FEBID process,
electron-beam induced fragmentation and reactivity on a substrate is a complex problem that combines many interlinked phenomena, such as the interaction of precursor molecules with a substrate, primary beam transport and electron multiple scattering on the substrate, ejection of backscattered and secondary electrons of different energies, their interaction with precursors by different mechanisms (e.g., DEA or DI) leading to different fragmentation pathways, reactivity of the fragmentation products among themselves and with other molecules, etc.
%
Mass spectrometry experiments with electron beams represent a convenient means to test and validate the molecular fragmentation models, since the abundance of fragments of different masses and charges can be accurately measured under single electron--molecule collision conditions.

The goal of this work is to develop and validate a model for molecular fragmentation after energy deposition that can be used in IDMD simulations of FEBID \cite{Sushko2016} and in other situations where IDC is relevant, e.g., for studying radiation-induced biodamage \cite{Nano-IBCT_book, deVera_2018_EPJD.72.147}. 
As an illustrative case study we analyze fragmentation of a tungsten hexacarbonyl,  W(CO)$_6$, molecule for which mass spectrometry data on electron impact ionization is abundant \cite{Wnorowski_2012_IJMS.314.42}. It is demonstrated that the quantitatively correct fragmentation picture including different fragmentation channels can be reproduced by means of classical MD simulations. 
We show that loss of CO ligands is mainly the result of the thermal evaporation process wherein excess energy is distributed among all degrees of freedom of the parent molecule. This mechanism cannot describe the formation of WC(CO)$_n^+$ fragments, which are produced due to the initial cleavage of a C--O bond. This process occurs due to localized energy deposition in the particular bond. Calculated appearance energies for the production of different fragments are in good agreement with experimental data \cite{Wnorowski_2012_IJMS.314.42}.

\section{Computational methodology}
\label{sec:methods}

MD simulations were performed in this work using MBN Explorer \cite{Solovyov2012} -- a software package for advanced multiscale modeling of complex molecular structure and dynamics. Among many options for molecular dynamics available in the software, it includes the reactive CHARMM force field \cite{Sushko2015} which was utilized in this work as well as the IDMD module \cite{Sushko2016} which will benefit from the fragmentation model developed in this study. The program counts on with the MBN Studio graphical interface \cite{MBNStudio_paper_2019} which was used to construct the molecular system, perform the simulations and analyze the results. In the following sections parameters of the reactive force field and the model of W(CO)$_6$ fragmentation are described in detail.

\subsection{Reactive CHARMM force field}
\label{sec:reactiveFF}

\begin{sloppypar}
The reactive CHARMM force field \cite{Sushko2015} permits classical MD simulations of a large variety of molecular systems experiencing chemical transformations whilst monitoring their molecular composition and topology changes \cite{MBNbook_Springer2017, Sushko2016, Sushko2015}. It can be applied for studying processes where rupture of chemical bonds plays an essential role, e.g., in irradiation- or collision-induced damage \cite{Verkhovtsev_2017_EPJD.71.212, deVera_2018_EPJD.72.147}. 
\end{sloppypar}

This reactive force field represents a modification of the standard CHAR\-MM force field \cite{MacKerell1998}, which employs the harmonic approximation for the description of interatomic interactions thereby limiting its applicability to small deformations of molecular systems. 
Contrary to the standard CHARMM force field, its reactive modification goes beyond the harmonic approximation for modeling covalent bonds, angles and dihedral angles. This allows for a more accurate description of the physics of molecular dissociation. In our recent works the reactive CHARMM force field was utilized to study ion-induced radiochemistry \cite{deVera_2018_EPJD.72.147} and thermal splitting \cite{Sushko2015} of water as well as collision-induced multi-fragmentation of C$_{60}$ fullerenes \cite{Verkhovtsev_2017_EPJD.71.212}. Here it is used to study fragmentation of organometallic compounds.

In the standard CHARMM force field \cite{MacKerell1998} all bond\-ed interactions are described by harmonic potentials. For example, the interaction between atoms $i$ and $j$ forming a covalent bond is given by:
\begin{equation}
U^{\rm bond}(r_{ij}) = k_{ij}^{\rm bond}(r_{ij}-r_{0,ij})^2 \mbox{ , }
\label{eq:bondharm}
\end{equation}
where $k_{ij}^{\rm bond}$ is the force constant of the bond, $r_{ij}$ is interatomic distance and $r_{0,ij}$ is the equilibrium bond length. Angular and dihedral interactions involving triples and quadruples of atoms are defined in a similar manner \cite{MacKerell1998}. 

\begin{figure*}[htb!]
\centering
\includegraphics[width=0.98\textwidth]{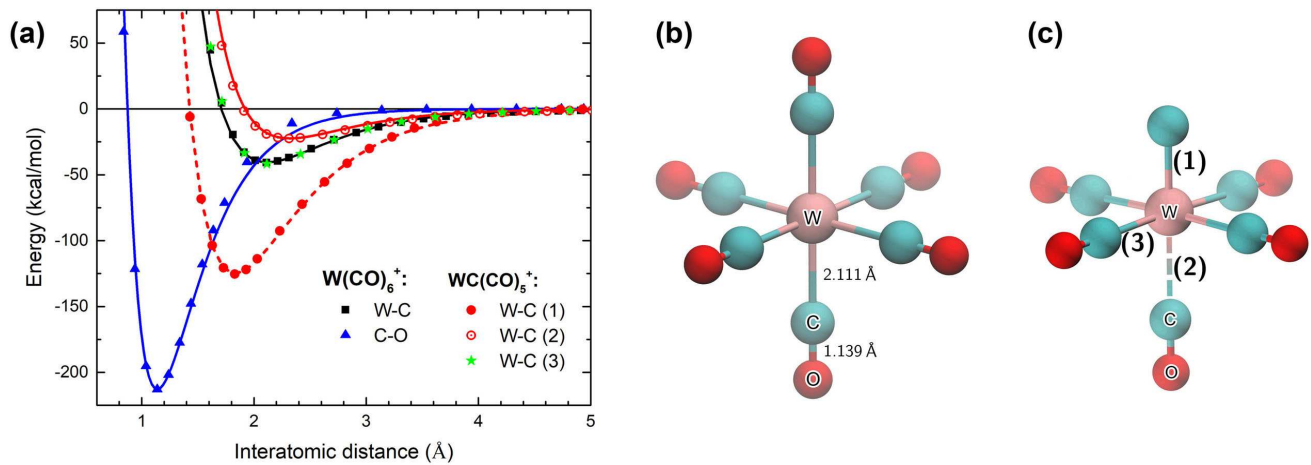}
\caption{
Panel (a): Potential energy curves for W--C and C--O bonds in the W(CO)$_6^+$ and WC(CO)$_5^+$ molecules. 
Symbols show the results of DFT calculations performed using the B3LYP functional and a LanL2DZ/6-31+G(d,p) basis set. 
Lines show the fit to these data with the Morse potential, Eq. (\ref{eq:Morse}).
Panels (b) and (c) show, respectively, the optimized structures of W(CO)$_6^+$ and WC(CO)$_5^+$. In the latter case, three different types of W--C bonds (labeled as (1), (2) and (3)) appear after one oxygen atom has been removed.
}
\label{fig_WCO6_PES}
\end{figure*}

In the reactive force field \cite{Sushko2015}, the harmonic bond interaction described by Eq.~(\ref{eq:bondharm}) is replaced by a Morse potential:
\begin{equation}
U^{\rm bond}(r_{ij}) = D_{ij}\left[ e^{-2\beta_{ij}(r_{ij}-r_{0,ij})}-2e^{-\beta_{ij}(r_{ij}-r_{0,ij})} \right] \mbox{ . }
\label{eq:Morse}
\end{equation}
Here $D_{ij}$ is dissociation energy for the bond between atoms $i$ and $j$, and the parameter $\beta_{ij}=\sqrt{k_{ij}^{\rm bond}/D_{ij}}$ determines steepness of the potential. An additional parameter can be defined, the cutoff distance, characterizing the distance beyond which the bond is considered as broken and the molecular topology of the system changes.
The bond energy calculated by Eq. (\ref{eq:Morse}) asymptotically approaches zero at large interatomic distances. Switching functions are defined accordingly to gradually reduce the angular and dihedral interactions as the bond breaks \cite{Sushko2015}.

\subsection{Determination of the force field parameters}
\label{sec:parameters}

Parameters of the reactive force field for W(CO)$_6$ and its fragments 
were determined from DFT calculations which were benchmarked against experimental data \cite{Wnorowski_2012_IJMS.314.42}. The DFT calculations were performed using Gaussian 09 software \cite{Gaussian09} employing the B3LYP exchange-correlation functional and a mixed LanL2DZ/6-31+G(d,p) basis set, whe\-rein the former set described the W atom and the latter was applied to C and O atoms. 
As we aim at reproducing appearance energies from electron-impact ionization experiments, singly charged parent molecule and molecular fragments were considered. 
Geometry of each molecule was optimized first (states with spin multiplicity $M = 2, 4$ and 6 were considered) and a potential energy surface scan was then performed for different W--C and C--O bonds allowing to calculate equilibrium bond lengths, dissociation energies and force constants. Atomic partial charges were obtained through the natural bond orbital analysis \cite{Gaussian09}.

\begin{table*}
\centering
\caption{Parameters of the reactive force field used in the simulations. In the case of WC(CO)$_n^+$, W--C bonds labeled as (1) and (2) refer to the notations of Fig.~\ref{fig_WCO6_PES}(c).}
\begin{tabular}{p{2.8cm}p{1.3cm}p{1.3cm}p{1.3cm}p{1.3cm}p{1.3cm}}
\hline 
           & \multicolumn{2}{c}{W(CO)$_n^+$} & \multicolumn{3}{c}{WC(CO)$_n^+$} \\ 
\hline
bond type      &   W--C     &   C--O         &  W--C (1)  & W--C (2) & C--O   \\
\hline
$r_0$~(\AA)      &    2.11     &     1.14    &  1.83      &  2.31   &  1.14  \\
$D$~(kcal/mol) &    40.7     &    212.8    &  143.3     &  22.6   &  212.8  \\
$k^{\rm bond}$(kcal/mol/\AA$^2$) & 119.9 &   1493.9   &  369.4     &  67.7   &  1493.9 \\
\hline
\end{tabular}
\label{table:CHARMM_parameters}
\end{table*}


To benchmark the methodology, we analyzed equilibrium bond lengths in a neutral W(CO)$_6$ molecule. The calculated values, $r_{0,{\rm{W-C}}} = 2.07$~\AA, $r_{0,\rm{C-O}} = 1.15$~\AA, are in good agreement with experimental data \cite{Arnesen_1966_ActaChemScand.20.2711} and with the results of earlier DFT calculations \cite{Szilagyi_1997_Organometallics.16.4807}. 
The calculated dissociation energies of W--C bond in the parent cation W(CO)$_6^+$ and in different fragments such as W(CO)$_n^+$ ($n= 1-5$) and WC(CO)$_n^+$ ($n = 0-5$) are in good agreement with the results of Ref.~\cite{Wnorowski_2012_IJMS.314.42} with the relative discrepancy of a few kilocalories per mole.

Figure~\ref{fig_WCO6_PES}(a) shows potential energy curves for W--C and C--O bonds in the parent W(CO)$_6^+$ molecule as well as in WC(CO)$_5^+$ formed upon removal of an oxygen atom from one of the ligands. Optimized structures of these molecules are shown in panels (b) and (c), respectively. Symbols illustrate the results of DFT calculations while lines show a fit to this data with the Morse potential, Eq. (\ref{eq:Morse}). The W--C bond in the parent molecule is significantly weaker ($D = 40.7$~kcal/mol) than the C--O bond ($D = 212.8$~kcal/mol), which is a common feature of metal carbonyls \cite{Diefenbach_2000_JACS.122.6449}.  
In the case of WC(CO)$_5^+$, three different types of W--C bonds, labeled as (1), (2) and (3), can be distinguished, see Fig.~\ref{fig_WCO6_PES}(c). When an oxygen atom is removed, the remaining carbon atom of the ligand becomes stronger bound to the tungsten atom (bond (1)); the dissociation energy of this W--C bond increases and varies from 119.9 to 143.3 kcal/mol depending on the WC(CO)$_n^+$ fragment considered. The opposite CO group becomes weakly bound to the metal atom (W--C bond (2)) and the equilibrium distance between W and C increases from 2.11~\AA~to 2.31~\AA. Four W--C bonds in the perpendicular direction (bond (3)) remain unaffected and their dissociation energy does not change with respect to that in the parent molecule.

Table~\ref{table:CHARMM_parameters} summarizes the parameters of the reactive force field used in the calculations. 
Since the dissociation energy of W--C bond in different W(CO)$_n^+$ ($n= 1-5$) fragments does not vary significantly~\cite{Wnorowski_2012_IJMS.314.42}, the value for the formation of W(CO)$_5^+$, 
$D = 40.7$~kcal/mol, was used in the simulations.

\subsection{Model for energy deposition}
\label{sec:Edepos}

Molecular fragmentation as a result of energy deposition involves several stages that take place on different time scales.
First, the incident radiation (e.g., an electron) interacts with the molecule and transfers energy to it by means of different mechanisms, e.g., electronic excitation, ionization or DEA. These are fast processes that happen on the sub-femtosecond scale and leave the molecule in an excited electronic state. In the case of ionization, some fraction of deposited energy is spent in overcoming the ionization threshold, another fraction is carried away by the ejected electron, while the remaining part is stored in the target in the form of electronic excitations. The latter can involve different molecular orbitals, being of either bonding or antibonding nature.

An excitation involving an antibonding molecular orbital will quickly evolve through cleavage of a particular bond on the femtosecond timescale. The fragmented parent molecule may still keep some amount of the deposited energy which can lead to the sequential fragmentation of other bonds on the picosecond or even longer timescales.

The excited electronic state may also involve a bonding molecular orbital or may not be localized on a particular bond. In this case, the excess energy can be redistributed over a larger part or even the entire volume of the system and be transferred later into its vibrational degrees of freedom.
Relaxation of the deposited energy due to electron-phonon coupling mechanism \cite{Gerchikov_2000_JPB.33.4905} leads to an increase in the amplitude of thermal vibrations which, in turn, leads to evaporation of loosely bound CO ligands. As it was shown in the case of small metal clusters  \cite{Gerchikov_2000_JPB.33.4905}, the electron-phonon coupling is a slow process (as compared to a characteristic time of electron-molecule interaction) that happens on a picosecond time scale. The subsequent evaporation process may last up to microseconds.

\begin{figure*}[htb!]
\centering
\includegraphics[width=0.65\textwidth]{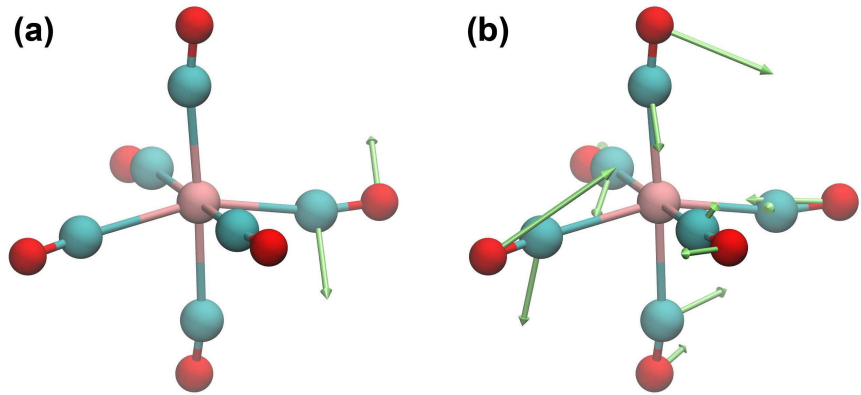}
\caption{
Schematic illustration of the two stages of fragmentation of W(CO)$_6$. (a) Stage I: the deposited energy is transferred into kinetic energy of atoms in a specific bond. The resulting velocities of the atoms are shown by arrows. (b) Stage II: the deposited energy is redistributed over the entire molecule and shared between the kinetic energy of all the atoms.
}
\label{fig:fragm_models}
\end{figure*}

In this work, we focus on the events happening from the cleavage of a particular bond to the redistribution of excess energy over the internal degrees of freedom of the molecule, i.e., on the time scales spanning from pico- to microseconds. Although quantum mechanical calculations can simulate the process of energy deposition, excitation of specific electronic orbitals and the initial stage of the fragmentation process, the span of the entire fragmentation process goes far beyond the limits of quantum MD. Thus, classical MD remains the only computational  technique allowing exploration of the process within the required time frame at the atomistic level of detail.
Quantum mechanical calculations would be useful though to establish the initial conditions for the multiscale simulation of molecular fragmentation, but this analysis goes beyond the scope of the present paper.

Within the framework of classical MD, we propose the following approach to describe the aforementioned fragmentation stages, i.e., fast cleavage of an individual bond (referred hereafter as stage I) and slow energy redistribution over all the molecular degrees of freedom (stage II from now on).
Both processes result in an increase of internal energy of the molecule after the energy deposition, which is treated as an initial increase of the kinetic energy of atoms.

Within the approach we have used to model stage I the energy is deposited locally into a specific covalent bond of the target and converted into kinetic energy of the two atoms forming the bond (see Fig.~\ref{fig:fragm_models}(a)). 
Velocities of these atoms are defined to obey the total energy and momentum conservation laws:
\begin{equation}
\vec{v}_1 =  \frac{ \sqrt{2\mu E_{\rm dep} } }{m_1}\vec{u} \ , \qquad
\vec{v}_2  =  -\frac{ \sqrt{2\mu E_{\rm dep} } }{m_2}\vec{u} \ . 
\label{eq:modtwo}
\end{equation}
Here $E_{\rm dep}$ is the amount of deposited energy remaining in the system after ionization (i.e., excess energy over the first ionization potential), $m_1$, $m_2$ and $\mu = m_1 m_2 / (m_1 + m_2)$ are, respectively, masses and the reduced mass of the atoms forming the bond, and $\vec{u}$ is a unit vector defining the direction of the relative velocity of these atoms upon bond cleavage. The orientation of $\vec{u}$ may be determined by the field resulting from the local molecular configuration around the bond. In the current model, it has been chosen randomly.

Stage II is governed by the thermal mechanism of fragmentation where the energy is distributed over all degrees of freedom of the target. 
In this case, equilibrium velocities of atoms corresponding to a given temperature, $v_{i}^{\rm eq}$, 
are scaled\footnote{Note that only the absolute values of the velocities are scaled, while their directions are unaltered; thus, the total momentum is conserved.} by a factor $\alpha$ depending on the amount of energy deposited (see Fig.~\ref{fig:fragm_models}(b)). The kinetic energy of $N$ atoms is then given by:
\begin{equation}
\sum_i^{N} \frac{1}{2}m_i (\alpha \, v_i^{\rm eq})^2 =  \frac{3N k_{\rm B}T}{2}+ E_{\rm dep} \ , 
\label{eq:modall}
\end{equation}
where the first term on the right hand side corresponds to the initial kinetic energy of the atoms at equilibrium (e.g., $T = 300$~K in our simulations), with $k_{\rm B}$ being the Boltzmann's constant, while the second term is the excess energy deposited in the molecule during the collision.

As it will be discussed in Section~\ref{sec:results}, 
each of the above-described mechanisms leads to
the formation of a particular group of experimentally observed fragments, while the whole experimental picture is reproduced well when both stages of the fragmentation process are considered.
One should note that these stages happen sequentially in the same excited molecule, so that the initial cleavage of a specific bond may be followed by redistribution of the remaining energy over the molecular fragment.

\subsection{Simulation details}
\label{sec:simulations}

\begin{sloppypar}
The MD simulations of W(CO)$_6$ fragmentation were performed using the MBN Explorer software \cite{Solovyov2012}. First, structure of the molecule was optimized using the parameters obtained from the DFT calculations discussed above. Then the molecule was equilibrated at $T = 300$~K for 100~ns. The equilibration simulation was performed using the Langevin thermostat with damping time of 2~ps. Atomic coordinates and velocities were recorded every 100~ps.
The equilibrated trajectory was sampled to generate random initial geometries and velocity distributions for the simulation of fragmentation. About 1000 constant-energy simulations each of 1~$\mu$s duration were conducted at different values of $E_{\rm dep}$ ranging from 0 to 475 kcal/mol. The upper limit is several times larger than the energy needed to break one W--C bond (see Fig.~\ref{fig_WCO6_PES}) which enables simulation of evaporation of several CO ligands. Fragments produced after 1~$\mu$s of simulation were analyzed, and the corresponding appearance energies were evaluated from this analysis and compared to experimental data.
\end{sloppypar}

\section{Results and discussion}
\label{sec:results}

In the following, the fragmentation pathways of a W(CO)$_6$ molecule upon electron impact ionization are discussed using the energy deposition model described in Section~\ref{sec:Edepos}. Simulated appearance energies are compared to the most recent set of experimental data obtained from mass spectrometry for positively charged fragments produced by electron beams of energy $\leq 140$ eV  \cite{Wnorowski_2012_IJMS.314.42}, as well as to the previously reported experimental values \cite{Michels_1980_InorgChem.19.479, Winters_1965_InorgChem.4.157, Bidinosti_1967_CanJChem.45.641, Foffani_1965_ZPhysChem.45.79, Qi_1997_JCP.107.10391}.

\begin{figure}[t]
\centering
\includegraphics[width=1.0\columnwidth]{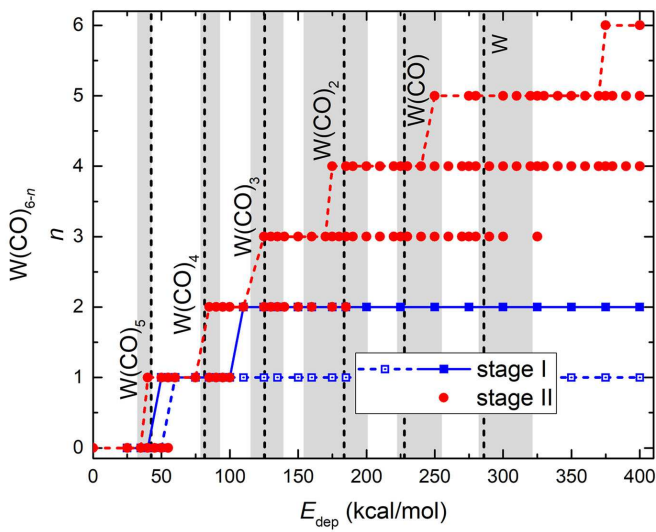}
\caption{Appearance energies for $n$ CO ligands produced upon fragmentation of a W(CO)$_6$ molecule. Results of the reactive MD simulations performed in this work are shown by symbols (see the text for details). Experimental appearance energies from Ref.~\cite{Wnorowski_2012_IJMS.314.42} are shown by black dashed lines and from Refs. \cite{Michels_1980_InorgChem.19.479, Winters_1965_InorgChem.4.157, Bidinosti_1967_CanJChem.45.641, Foffani_1965_ZPhysChem.45.79, Qi_1997_JCP.107.10391} by shaded areas. 
Dashed red line shows the largest number of CO released at each $E_{\rm dep}$.
}
\label{fig:AEWCO6}
\end{figure}

Figure~\ref{fig:AEWCO6} shows the number of CO fragments, $n$, produced upon dissociation 
of the parent W(CO)$_6^+$ molecule for a given amount of excess energy $E_{\rm dep}$. 
Open blue squares show the results of simulations of stage~I.
In this case, a given amount of energy was deposited into a W--C bond resulting in a prompt release of one CO group. However, no further 
fragmentation has been observed 
even at high values of $E_{\rm dep}$.
Due to a large difference in masses of a carbon and a tungsten atoms more than 90\% of deposited energy is transferred into kinetic energy of the C atom and carried away by the released CO group. Therefore, independently of the value of $E_{\rm dep}$ given to the molecule through cleavage of a W--C bond, only a small amount of energy is transferred to the remaining W(CO)$_5$ fragment, which is not sufficient to observe further fragmentation events. 

As discussed in more detail further in this section, a single atom or a small fragment produced after cleavage of a specific bond can hit the remaining large fragment upon its escape, redepositing some amount of energy into the large fragment and thus triggering further fragmentation at stage II.
Filled blue squares describe the situation when the CO group released due to cleavage of a W--C bond collided with the remaining W(CO)$_5$ molecule, which led to the loss of another CO group. However, not more than two CO ligands have escaped the parent molecule in this case.

The results of simulations describing stage II
are shown in Fig.~\ref{fig:AEWCO6} by filled red circles. 
Distribution of deposited energy over all degrees of freedom of the target leads to evaporation of multiple CO fragments. 
These results
are in good agreement with the appearance energies reported in a recent experimental study by Wnorowski \textit{et al.}~\cite{Wnorowski_2012_IJMS.314.42} (black dashed lines) as well as with appearance energies obtained in earlier experiments \cite{Michels_1980_InorgChem.19.479, Winters_1965_InorgChem.4.157, Bidinosti_1967_CanJChem.45.641, Foffani_1965_ZPhysChem.45.79, Qi_1997_JCP.107.10391} (shaded areas). Note that the first ionization potential was subtracted from the experimental appearance energies to convert them into the excess deposited energy.  
Due to statistical nature of the fragmentation process, emission of a given number of CO groups was observed at different values of $E_{\rm dep}$. 
The lowest values at which five or less CO molecules were recorded after 1~$\mu$s agree nicely with the experimental results. However, complete fragmentation (i.e, loss of 6 CO molecules) has been observed only in a few trajectories at $E_{\rm dep} = 375$ and 400~kcal/mol, which are significantly higher than the experimental appearance energy for W$^+$. 
This indicates that the complete fragmentation takes place on a larger time scale and longer simulations are needed to observe the complete fragmentation at smaller energies.

\begin{figure}[t]
\centering
\includegraphics[width=1.0\columnwidth]{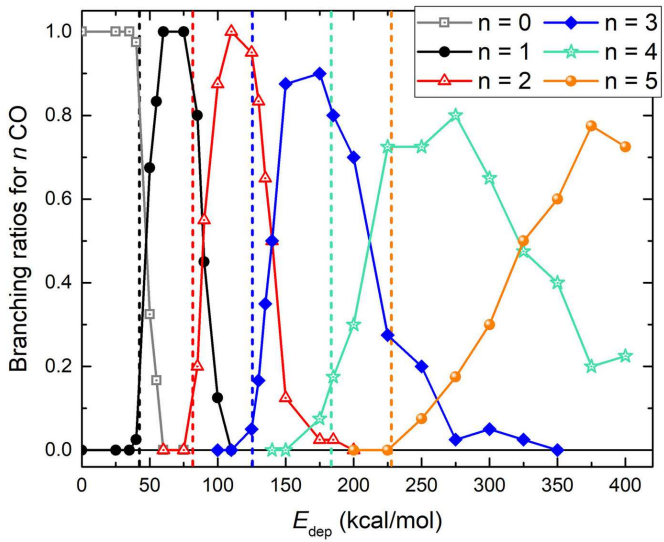}
\caption{Branching ratios for the formation of $n$ CO ligands upon fragmentation of a W(CO)$_6$ molecule during stage II as a function of the excess energy $E_{\rm dep}$. At a given $E_{\rm dep}$, the sum of branching ratios for all the fragments is equal to 1. Results of the reactive MD simulations performed in this work are shown by symbols. Experimental appearance energies from Ref.~\cite{Wnorowski_2012_IJMS.314.42} are shown by dashed lines.}
\label{fig:AEWCO6_prob}
\end{figure}

The results of simulations describing stage II
can be used to evaluate branching ratios for the production of different fragments for a given amount of excess energy $E_{\rm dep}$. This analysis provides information on how many carbonyl groups will be most likely evaporated at a given $E_{\rm dep}$ after 1 $\mu$s. These results are presented in Figure~\ref{fig:AEWCO6_prob}. It shows that emission of one, two and three CO groups ($n = 1, 2, 3$) takes place in rather well separated energy ``windows''. 
Fragments corresponding to $n=1$ and 2 were recorded in these energy ranges with the maximal probability corresponding to the branching ratio of 1. 
For the larger numbers of emitted CO the maximal branching ratios drop down to 0.8 suggesting an increased probability of observing different fragments. Note also that the characteristic energy ranges for the emission of $n$ CO groups increase with $n$. 
This analysis allows for evaluation of a typical amount of energy that should be deposited into the W(CO)$_6$ molecule to observe a specific fragment.
These values will be used in our future work as input for IDMD simulations in order to define the probabilities for different fragmentation pathways.

\begin{figure}[t]
\centering
\includegraphics[width=1.0\columnwidth]{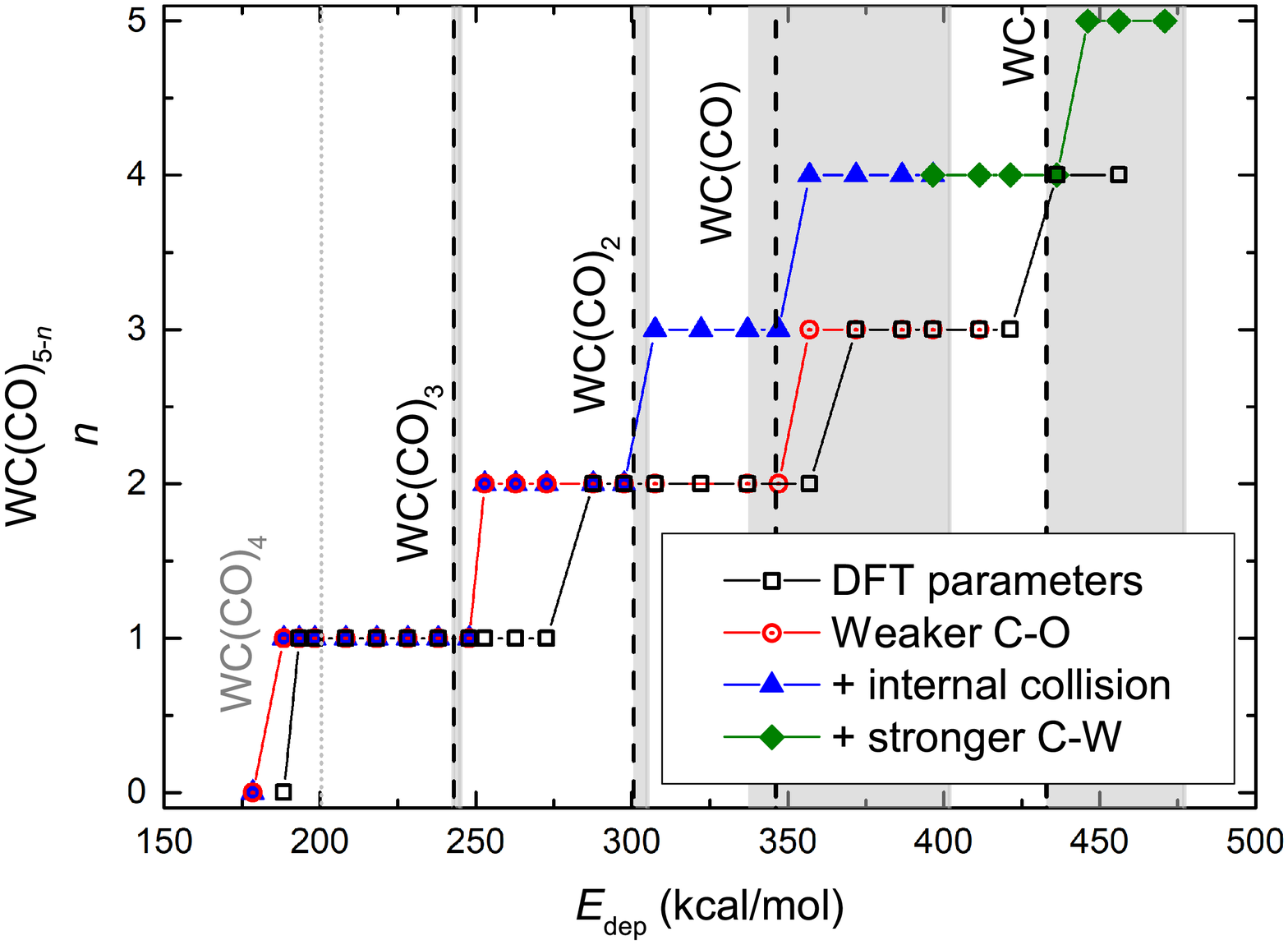}
\caption{Comparison between simulated (symbols) and experimental appearance energies for the different fragments of the WC(CO)$_5^+$ molecule.
Open black and red symbols correspond to simulations in which the oxygen atom escapes the system without further collision (see the text for details). Filled blue and green symbols describe events where the O atom collides with the parent molecule. Experimental appearance energies from Ref.~\cite{Wnorowski_2012_IJMS.314.42} are shown by dashed lines and from Refs. \cite{Michels_1980_InorgChem.19.479, Winters_1965_InorgChem.4.157, Bidinosti_1967_CanJChem.45.641, Foffani_1965_ZPhysChem.45.79, Qi_1997_JCP.107.10391} by shaded areas.}
\label{fig:AEWC(CO)5}
\end{figure}

The experimental fragmentation mass-spectra~\cite{Wnorowski_2012_IJMS.314.42, Michels_1980_InorgChem.19.479, Winters_1965_InorgChem.4.157, Bidinosti_1967_CanJChem.45.641, Foffani_1965_ZPhysChem.45.79, Qi_1997_JCP.107.10391} revealed the formation of not only W(CO)$_{6-n}^+$ ($n =  0- 6$) but also WC(CO)$_{5-n}^+$ ($n = 2-5$) molecules, 
which, however, have not been observed in the simulations of stage II
even after 1~$\mu$s of simulation. This is due to the very low probability for observing a statistical cleavage of a C--O bond, owing to the much lower strength of the W--C bond. 
We therefore assumed that a C--O bond can break after a localized energy deposition into it.
As described in Section~\ref{sec:parameters}, loss of an oxygen atom from a CO group makes the opposite ligand weaker bound to the W atom (bond (2)), while the W--C bond corresponding to the cleaved C--O (bond (1)) becomes much stronger. Therefore, localized energy deposition into one C--O bond leads to a prompt release of an O atom, together with the recoil of the C atom to the parent molecule, which ends up being vibrationally excited and subsequently releases a CO fragment. This scenario explains also the production of  smaller fragments due to subsequent loss of several CO groups from WC(CO)$_5$ which were observed experimentally and
in the simulations at $E_{\rm dep} > 180$~kcal/mol, as shown in Fig.~\ref{fig:AEWC(CO)5}.

\begin{figure*}[htb!]
\centering
\includegraphics[width=0.8\textwidth]{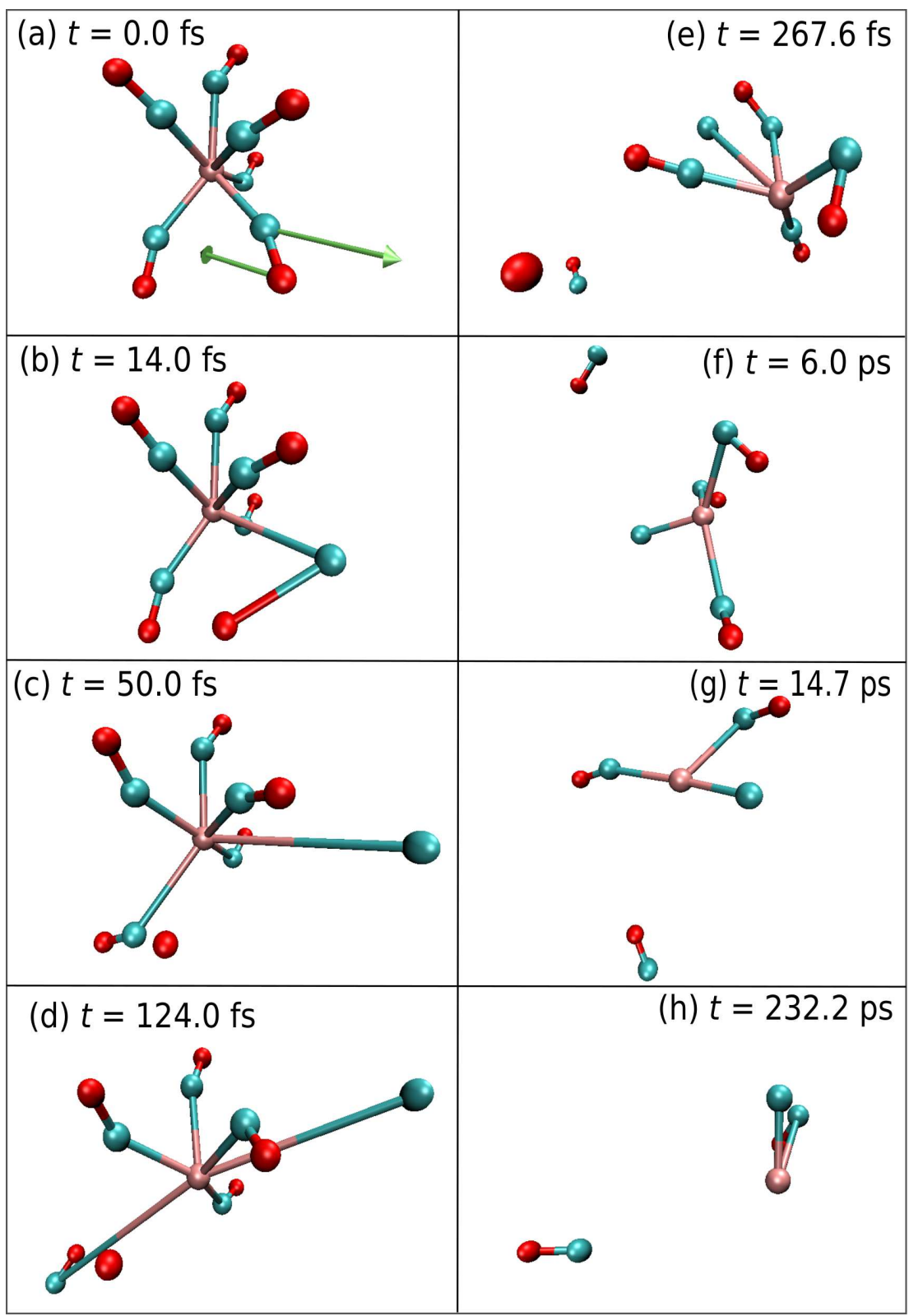}
\caption{
Snapshots of a trajectory of the fragmentation process through stages I and II, in which the escaping oxygen atom collides with a neighbouring CO group. The excess energy deposited into the C--O bond is 375 kcal/mol. The trajectory is centered on the W atom for clarity. See the text for further details.
}
\label{fig:trajectory}
\end{figure*}

The fragmentation mechanism in this case is more complex than the simple evaporation of CO molecules, and thus deserves a more detailed analysis. 
An exemplary simulated trajectory for $E_{\rm dep} = 375$~kcal/mol is shown in Fig.~\ref{fig:trajectory} (a movie is provided as supplementary material). Initial velocities of C and O atoms are shown by arrows in panel (a). As follows from the analysis of this trajectory, the initial cleavage of the C--O bond is a fast event happening within the first 50~fs of the simulation (panels (b-c)).
It leads to the release of an energetic O atom as a result of energy redistribution according to Eq.~(3) (stage I). This atom can directly escape from the parent molecule or collide with it in its way out, depending on the orientation of the vector $\vec{u}$. In the former case, less extensive fragmentation is observed as the O atom is ejected with a notable amount of kinetic energy. In the latter case, a fraction of its kinetic energy is redeposited into the parent molecule leading to further fragmentation (stage II).

The collision of the escaping O atom with the parent molecule is depicted in the next panels of Fig. \ref{fig:trajectory}.
After about 50~fs (panel (c)) the ejected O atom collides with a neighbouring CO ligand. The latter is ejected from the parent molecule after approx. 130~fs (panel (d)) while the remaining C atom, becoming strongly bound to W, does not detach from the parent molecule (panel (e)). The remaining molecular fragment is still excited, and the evaporation process follows with the ejection of a CO group by 6~ps (panel (f)), another one by $\sim 15$~ps (panel (g)) and the last one by $\sim 230$ ps (panel (h)).

It should be noted that in this case the reactive force field is rather sensitive to the parameters used, particularly to the dissociation energies, and variation of these parameters can impact the results of simulations. In Fig.~\ref{fig:AEWC(CO)5}, black squares depict the results of simulations using the dissociation energies calculated by DFT (see Section~\ref{sec:parameters}), specifically $D_{\rm C-O} = 212.8$ kcal/mol. While the appearance energy for WC(CO)$_4$ is well reproduced,\footnote{Although this fragment was not observed experimentally, it appearance energy (gray dotted line in Fig.~\ref{fig:AEWC(CO)5}) can be estimated from the data reported in Ref.~\cite{Wnorowski_2012_IJMS.314.42}. The calculated first appearance energy is smaller than the expected experimental value: this is due to the thermal energy of the molecule which also contributes to fragmentation together with the deposited energy.} the rest are shifted towards larger energies. However, by analyzing the appearance energies reported by Wnorowski \textit{et al.} \cite{Wnorowski_2012_IJMS.314.42}, one can estimate the dissociation energy of C--O bond of about 180~kcal/mol. It should be noted that the C--O bond in metal carbonyls is rather complex and can be characterized as a mixture of a triple and a double bond; $D_{\rm{C-O}}$ varies from 126 kcal/mol for a double bond to 257 kcal/mol for a triple bond \cite{Darwent1970,Dean1972}. Thus, values within this range are meaningful. 
Simulation results using $D_{\rm{C-O}} = 180$~kcal/mol are shown in Fig.~\ref{fig:AEWC(CO)5} by open red circles. The calculated appearance energies are shifted to lower energies with respect to the results employing the dissociation energy obtained from DFT, being closer to the experimental results. 

Filled blue triangles show the results of simulations where the collision of the escaping O atom with the parent molecule has been taken into account. In this case, the first four appearance energies (WC(CO)$_4$ to WC(CO)) are well reproduced, but the last one (WC) is not observed. This happens because the deposition of large amounts of $E_{\rm dep}$ larger than 390 kcal/mol leads to the cleavage of not only C--O bond but also the W--C bond labelled as (1) in Fig. \ref{fig_WCO6_PES}(c).
One may expect that this bond becomes stronger as the coordination number of W decreases (i.e. when smaller fragments are formed) that happens at large values of $E_{\rm dep}$. Thus, $D_{\rm{W-C}}$ could be larger than 143~kcal/mol, the value obtained from the DFT calculations for the WC(CO)$_5^+$ molecule. We have therefore performed a set of simulations using $D_{\rm{W-C}} = 180$ kcal/mol, that is similar to the dissociation energy of the C--O bond. These results are shown by filled green diamonds in Fig.~\ref{fig:AEWC(CO)5}.

\begin{sloppypar}
Under these conditions, the experimental appearance energies are well reproduced. Although the reactive force field is rather sensitive to the parameters used, an appropriate choice of the parameters leads to a quantitative agreement with experiments. This shows that reactive classical MD simulations are appropriate to simulate the fragmentation patterns of W(CO)$_6$. The method, being general, can be applied to other organometallic precursors for FEBID as well as to many other inorganic, organic and biological molecules. 
\end{sloppypar}

\section{Conclusion and outlook}

We have developed and successfully validated a model for molecular fragmentation upon interaction with ionizing radiation. Two stages of the fragmentation process were considered, namely a localized energy deposition into a specific covalent bond and the redistribution of deposited energy among all the bonds of a target molecule. These situations represent limiting cases of the possible pathways for energy redistribution in a molecule, which involve the excitation to specific antibonding electronic orbitals or the energy transfer to vibrational modes of the molecule.

As a case study we analyzed fragmentation of a tungsten hexacarbonyl,  W(CO)$_6$, molecule which is widely used as a precursor for focused electron beam induced deposition (FEBID) and for which mass spectrometry data on electron impact ionization are abundant. We demonstrated that the quantitatively correct fragmentation picture including different fragmentation channels is reproduced well by means of classical MD simulations with reactive force fields.

The analysis revealed that loss of CO ligands and the formation of W(CO)$_n^+$ ($n = 0-5$) fragments is mainly the result of the thermal evaporation process where excess energy is distributed among all degrees of freedom of the parent molecule. Another type of fragments, WC(CO)$_n^+$ ($n = 0-4$), is produced mainly due to cleavage of a C--O bond as a result of the localized energy, followed by redistribution of excess energy in the parent molecule resulting in further evaporation of CO ligands. Calculated appearance energies for different fragments are in good agreement with experimental data \cite{Wnorowski_2012_IJMS.314.42}. 
We also analyzed the probability for the formation of different W(CO)$_n^+$ fragments as a function of deposited energy and evaluated characteristic values of deposited energy needed to release a given number of CO ligands. This information will be used in our future work to model FEBID of metal carbonyl precursors and the electron-induced formation and growth of metal nanostructures within the Irradiation Driven Molecular Dynamics approach \cite{Sushko2016}. 

\begin{sloppypar}
The fragmentation model considered in this study has a clear physical explanation and describes the different stages of electron-induced fragmentation of a molecular target. This approach can therefore be utilized to study radiation-induced fragmentation of other FEBID precursors (other metal carbonyls in particular) as well as biomo\-lecular systems whose fragmentation pathways are of great relevance for radiotherapy applications.
Results of this analysis may bring insights into irradiation-driven chemistry of various molecular systems which is exploited in different modern and emerging technologies. Additionally, the model developed represents a first step towards the modelling of fragmentation mass spectra by means of classical MD simulations. This objective, which goes beyond the scope of the present paper and might be accomplished in subsequent works, would require the development of appropriate quantum mechanical models for predicting the probabilities of energy deposition after an inelastic collision, as well as the probability of energy transfer through the localized or thermal mechanisms.
\end{sloppypar}

\section*{Acknowledgements}

PdV acknowledges the Alexander von Humboldt Foundation/Stiftung for its financial support by means of a postdoctoral fellowship. AV is grateful for the financial support from the DKFZ Postdoctoral Fellowship. 
This work was also supported in part by the Deutsche Forschungsgemeinschaft (Project no. 415716638), the Spanish Ministerio de Ciencia, Innovaci\'{o}n y Universidades and the European Regional Development Fund (Project no. PGC2018-096788-B-I00) and by the COST Action CA17126 ``Towards understanding and modelling intense electronic excitation'' (TUMIEE). 
The possibility to perform computer simulations at Goethe-HLR and FUCHS clusters of the Frankfurt Center for Scientific Computing is gratefully acknowledged.

\bibliographystyle{plain}
\bibliography{library}

\end{document}